# Extending the code generation capabilities of the Together CASE tool to support Data Definition Languages

M. Marino
*LBNL, Berkeley, CA 94720, USA*

Together is the recommended software development tool in the Atlas collaboration. The programmatic API, which provides the capability to use and augment Together's internal functionality, is comprised of three major components - IDE, RWI and SCI. IDE is a read-only interface used to generate custom outputs based on the information contained in a Together model. RWI allows to both extract and write information to a Together model. SCI is the Source Code Interface, as the name implies it allows to work at the level of the source code. Together is extended by writing modules (java classes) extensively making use of the relevant API. We exploited Together extensibility to add support for the Atlas Dictionary Language. ADL is an extended subset of OMG IDL. The implemented module (ADLModule) makes Together to support ADL keywords, enables options and generate ADL object descriptions directly from UML Class diagrams. The module thoroughly accesses a Together reverse engineered C++ project - and/or design only class diagrams - and it is general enough to allow for possibly additional HEP-specific Together tool tailoring.

## 1. INTRODUCTION

The ATLAS (see atlas.web.cern.ch/Atlas) Dictionary Language (ADL) [1] is a platform independent extended proper subset of the Interface Definition Language (IDL 2.0) from Object Management Group (see www.omg.org). ADL was developed as part of ATLAS continuing efforts to enhance and customize its software architecture and to provide support for object description and integration in Athena [2], the ATLAS off-line analysis framework.

The ADL description of data objects is used in the context of a general Data Dictionary (DD) facility and constitutes the input to compiler-based utilities (DD).

The utilities provided by the DD and based on ADL description [3] of objects can be as diverse as possible: data tools integration, semi-automatic persistency, schema evolution, multi-language support, component independence, stability and robustness, coding rules enforcement, etc.

In the late phase of ADL development one of the "would be nice if" envisaged evolution was the integration of ADL with Together CASE tool. The integration would have been an ideal solution to achieve the semi-automatic description of Even Data Model objects already developed and maintained within the collaboration. What was initially just a pure speculation became a reality after initial study of the open API of Together.

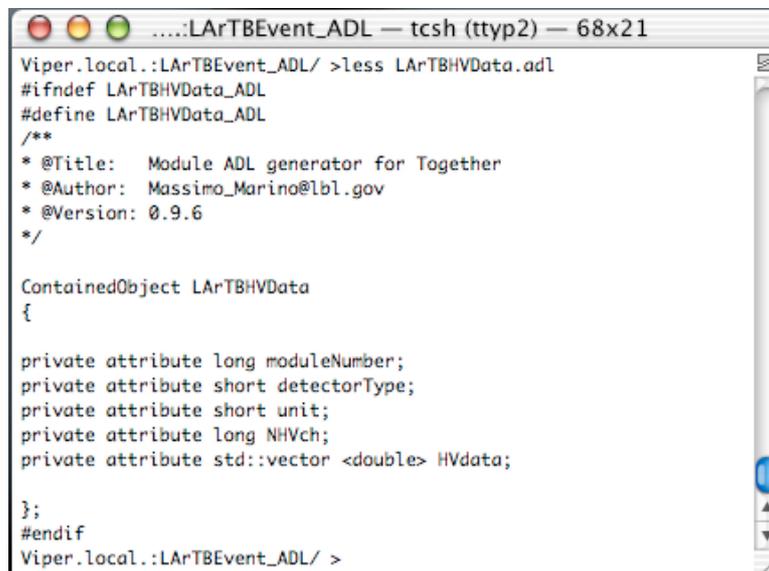

Figure 1: ADL description example

**Insert PSN Here**

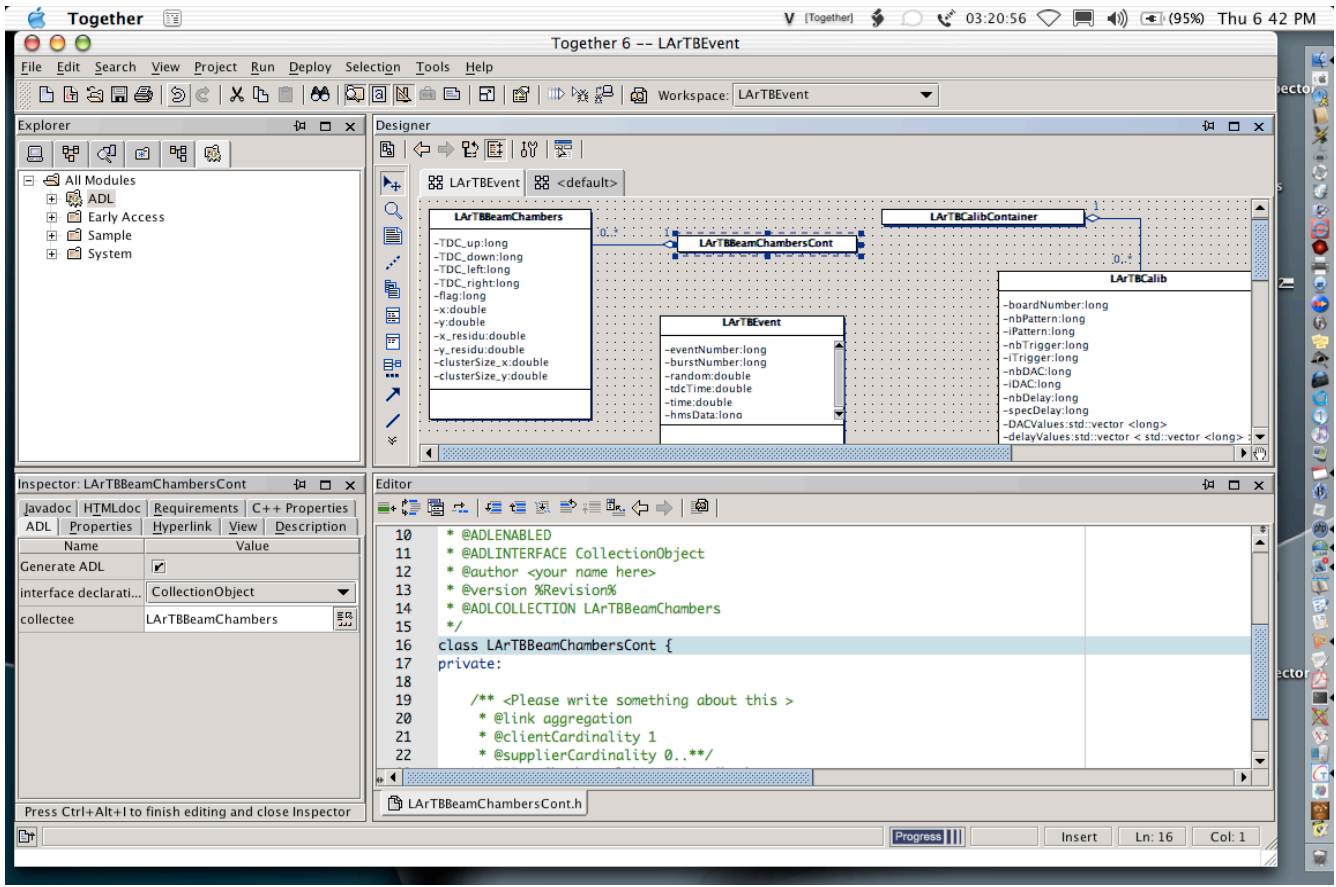

Figure 2: The ADL-Together integration

## 1.1. Together

Together [4] Control Center constitutes a complete environment covering the various aspects of the software development process: analysis, design, implementation and deployment. Together encompasses UML modeling with an integrated development environment (IDE) achieving a full synchronization between design and implementation. A characteristic of Together tool is its integrated dual view of the model: object oriented diagrams and the actual code constitute essentially a view of the same modeled information. Changes in the design can be reflected in the code nearly in a real time basis. This synchronization also occurs in the code editor so that changes in the implementation are automatically reflected in the model diagrams.

Together supports a wide range of platforms: Windows 95/98/NT/2000/XP, Sun Solaris, Linux, HP-UX, and Compaq Tru 64. Borland Enterprise Studio for Java supports Windows NT/2000/XP, Solaris, Linux, and Mac OS X. Together comes pre-integrated with common development tools for versioning, configuration management, requirements management, as well as an open API for integrating third party tools.

Together evaluation [5] took place within the Atlas collaboration in 2000 and it is the recommended lightweight CASE tool for development of C++ and Java programs [6].

## 2. THE OPEN API AND MODULES

## 2.1. Together Extensibility

Together Control Center comes with an open API composed of a three-tier interface that enables varying degrees of access to the native infrastructure. The top tier represents the highest degree of constraint and the lowest tier the least degree of constraint. The interfaces are very simply named:
      IDE
      Read-Write Interface (RWI)
      Source Code Interface (SCI

### 2.1.1. IDE

This is the API used to generate custom outputs based on information contained in a Together model. It is a read-only interface. IDE group provides the functionality related to the model representation in Together development environment and interaction with the user.

Each package composing the IDE group has a description highlighting its areas of applicability
      com.togethersoft.openapi.ide package and its sub-packages

**Insert PSN Here**

### 2.1.2. RWI

This API enables to go deeper into the Together architecture. One can both extract information from, and write information to the model and - to some extent = enhance Together capabilities.

RwiElement entities can represent more than packages, classes and members. In a RWI model they may represent different diagrams (class diagrams, use case diagrams, sequence diagrams and others), links, notes, use cases, actors, states, etc.

> com.togethersoft.openapi.rwi package and its sub-packages

### 2.1.3. SCI

As the name implies the Source Code Interface takes the developer down to the source code level. An SCI model is a set of sources (for Java, class files are allowed) organized into packages. The SCI packages represent the Java packages (which can be stored even in .zip or .jar files) or directories for other languages. SCI model can contain parts written in different languages.

SCI allows to work with the source code almost independently of the language being used. For example, a SciClass object can represent a class in both Java and C++.

> com.togethersoft.openapi.sci package and its sub-packages

### 2.1.4. Modules

Together comes with several pre-existing modules to extend its capability to integrate with other applications and tools. Modules are sets of Java classes that implement the IdeScript or the IdeStartup interface (or both) from the Together API. The ADL module makes use of the first two API groups.

## 3. THE ADL MODULE

This section will look at details of the ADL Generation module with code snips. Code snips will make explicit the exploitation of the open API and the relative little effort to achieve the desired – although visually complex – results.

The module development started with the InsertTags sample module in the Together tutorial – that is called plagiarism in literary circles, in the software industry it's called reuse.

The sample module provides the skeleton to traverse a Together model and add javadoc tags as it goes. It looks at classes, interfaces, and their members.

The ADL module builds on this and while traversing the Together project it extracts relevant information from selected classes in order to generates the ADL description file for each of them.
For this the module creates a new main Property
Inspector tab - simply labeled ADL - in the Together
GUI.

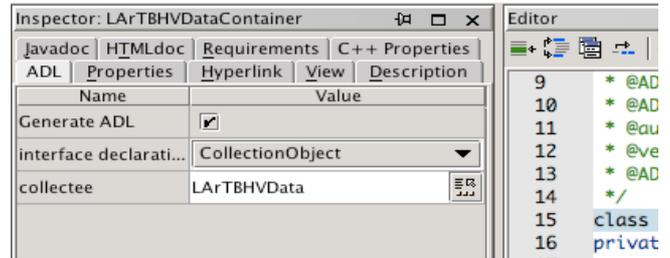

Figure 3: ADL tab in the Property Inspector

From the tab featured commands the user may select options on the ADL properties for the classes in the current diagram.

Selected options trigger the addition of relevant javadocs to the class declaration reflecting the selections made by the user on ADL properties.

The module operates into two phases: the first occurs at Together startup to configure the enhanced Property Inspector and read the configurations to properly run the ADL module, the second to gather the context of the opened Together project when the user selects classes in the current diagram and operate on the ADL options at class, method and attribute level as supported by the module itself.

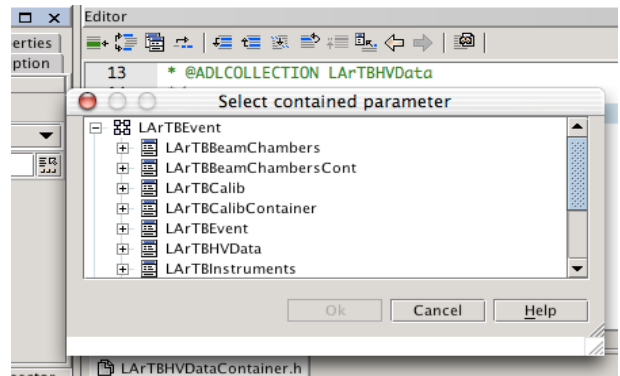

Figure 4: Select from classes in the model

### 3.1. Startup and Context Verification

There is not much in here apart instructing Together on where to look for the ADL module configuration file(s). Together will load any configuration file present in the config directory of the ADL module.

This allows for easy management as all different configurations can then be kept separated by their target, e.g., Property Inspector, diagrams, etc.

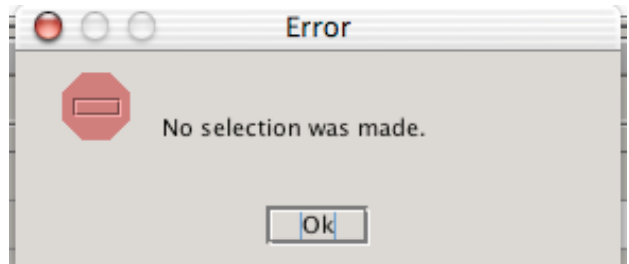

Figure 5: ADL module configuration example

When the ADL module is first invoked it makes sure that there is a project and diagram open, and issues a no action error message if there isn't:

```
if
(IdeProjectManagerAccess.getProjectManager().getActiveProject() == null)
{
IdeMessageManagerAccess.printMessage
(IdeMessageType.ERROR_MODAL, "No open project");
return;
}
IdeDiagramManager diagramManager =
IdeDiagramManagerAccess.getDiagramManager();
if (diagramManager.getActiveDiagram() == null)
{
IdeMessageManagerAccess.printMessage
(IdeMessageType.ERROR_MODAL, "No open diagram");
return;
}
```

Having assured that there is an opened project and diagram providing the module context, next task is to get the array of items (classes, methods, or attributes) that are selected. If nothing is selected, there is nothing to do and so the module returns with a no valid selection message:

```
RwiElement[ ] selectedRwiElements =
context.getRwiElements();
if (selectedRwiElements == null ||
selectedRwiElements.length == 0)
{
IdeMessageManagerAccess.printMessage
(IdeMessageType.ERROR_MODAL, "No selection was made.");
return;
}
```

Figure 6: Error message

If the module execution gets past this far there is a valid selection. Each item in the selection is then processed using a Visitor pattern [7]:

```
ADLOutVisitor adlVisitor = new ADLOutVisitor();
[…]
 if
(RwiShapeType.CLASS.equals(selectedRwiElement.getProperty(RwiProperty.SHAPE_TYPE))) {
//working with the selection
        selectedRwiElement.accept(adlVisitor);
}
```

### 3.2. ADLOutVisitor

This is the class that does the real work. It is an instance of the Visitor pattern. It recursively visits each element in the tree rooted at the current selection(s).

This class extends RwiVisitorAdapter, a class defined in the Together API specifically for writing model visitors.

Note in the previous code snip that elements have an accept(RwiVisitorAdaptor) method. The accept method then calls into the RwiVisitorAdaptor, passing itself as an argument. The method that is called depends on what the element being visited is.

The methods that ADLOutVisitor implements are:

```
public Object visitPackage(RwiPackage package){
return visitContainer(package); }
public Object visitDiagram(RwiDiagram diagram){
return visitContainer(diagram); }
public Object visitNode(RwiNode node){
return visitContainer(node); }
public Object visitMember(RwiMember member){
return visitElement(member); }
```

We need to provide methods for visiting each type of element we are interested in during the process of generating ADL descriptions.

#### 3.2.1. Visiting Packages
ADLModule assumption is that the user works on element of the project being opened, i.e., we do not want to work on imported elements or components. Making use of the concept of Property in the RWI API, we may ask questions about the identity and details of every element in the model.

**Insert PSN Here**

The set of properties may also be extended. This allows the ADLModule to define and set in the GUI – namely the Property Inspector – its own specific ones which are extensively used in the code.

In this particular case we check the value of the Together RWI property MODEL_PART so to verify that the selected package is indeed part of the project:

```
if (rwiPackage.hasProperty(RwiProperty.MODEL_PART)
```

Having ascertained that, we recursively visit each node in the supplied package:

```
while (rwiNodeEnumeration.hasMoreElements()) {
RwiNode nextRwiNode = rwiNodeEnumeration.nextRwiNode();

// Work only with enabled model elements
 if ("Model" != nextRwiNode.getProperty(RwiProperty.MODEL_PART)) {
visitNode(nextRwiNode);
    }
}
```

Next we visit every subpackages retrieving them with:

```
RwiPackageEnumeration subpackages = rwiPackage.subpackages();
```

Visiting diagrams is the same as visiting packages. We recursively visit the enclosed nodes & packages:

```
visitPackage(diagram.getContainingPackage());
```

Note that we do nothing to packages or diagrams being visited but get to their nodes, i.e., the singles classes therein described instead. The next method is where information relevant to ADL description generation is collected at class level.

### 3.2.2. Visiting Nodes

As before we only look at non-imported items, further we only want to look at class and interface nodes:

```
 if (RwiShapeType.CLASS.equals(rwiNode.getProperty(RwiProperty.SHAPE_TYPE))) {comment = (rwiNode.hasProperty(RwiProperty.INTERFACE)) ? "interface" : "class";
    // Get the writer of the ADL output file
 createAdlFileWriter(className);
```

At this point in the ADLModule we are inspecting a selected class. The createAdlFileWriter is responsible for attaching a new ADL file as <ClassName>.adl to an output streamer available to all methods in the module during the parsing of the details of each selected class:

**Insert PSN Here**

```
public void createAdlFileWriter(String className) {
try {
    // Open new ADL file
IdeMessageManagerAccess.printMessage (IdeMessageType.INFORMATION, "ADL generation for class "+ className);
File adlOutFile = new File(className + ".adl")
    // Attach Streamer
FileOutputStream adlOutStreamer = new FileOutputStream(adlOutFile);
    // Create ADL file writer
this.adlWriter = new PrintWriter(adlOutStreamer, true);
    }
catch (FileNotFoundException e) {
    System.err.println("*** ERROR(Text): can't handle adl file for class " + className + " " + e);
    }
  }
```

At this point we are in the process of parsing the class and writing the corresponding ADL declarations:

```
// Analyze Node
String adlInterface = rwiNode.getProperty("ADLINTERFACE");
// ADL Class declaratio
generateADLClassDeclaration(className, adlInterface, rwiNode);
// ADL includes
generateADLIncludes(className,adlInterface,rwiNode);
// Javadoc
[…]
```

We next visit in detail the class and each member. This includes both attributes & operations:

```
// Analyze members
RwiMemberEnumeration members = rwiNode.members();
while (members.hasMoreElements()) {
visitMember(members.nextRwiMember());
 }
```

### 3.2.3. Visiting Members

Method visitMember takes care of parsing attributes and methods for the selected class. Together API distinguishes them via the SHAPE_TYPE property:

```
if (RwiShapeType.OPERATION.equals (member.getProperty(RwiProperty.SHAPE_TYPE))) {
[…]
    visitOperation(member);
} else if (RwiShapeType.ATTRIBUTE.equals (member.getProperty(RwiProperty.SHAPE_TYPE))) {
[…]
    visitAttribute(member);
}
```

The visitAttribute method extracts the attribute NAME and TYPE. Then determines its visibility, thence the ADL persistent and read-only properties, thence it writes the information on the <classname>.adl file and repeats the process for each attribute. In the module these ADL specific properties are Booleans:

```
name = member.getProperty(RwiProperty.NAME);
t ype = member.getProperty(RwiProperty.TYPE);
[...]
// Determine attribute persistent/readonly proeprties
if (member.hasProperty("ADLREADONLY"))
adlAttrDecl = "readonly " + adlAttrDecl;
if (member.hasProperty("ADLPERSISTENT"))
adlAttrDecl = "persistent " + adlAttrDecl;
[...]
adlWriter.println(adlAttrDecl);
```

ADL properties selection from the Property Inspector are updated on the fly on the class source code with relevant Javadocs so to have immediately a visual clue of the selections being made from the GUI.

The GUI – Property Inspector – is independent of running the ADLModule: ADL properties selection are applied to the final ADL declaration output file when the module is actually run on the selected class(es).

The visitOperation method does several things to the javadocs of the class, subsequentely the parameters are processed for ADL generation. The return value of all methods is checked: An action is only valid if there is a return type  (i.e. constructors don't have one and do not have an ADL description) and it is not null:

```
if (member.hasProperty(RwiProperty.RETURN_TYPE))
```

In visitOperation we make use of RWIPropertyMap to pull the property values from each method parameter. At this point we have gathered and stored all required information to generate a meaningful ADL description for the class:

```
// Completed ADL
writeADL(className);
```

## Acknowledgments

This work would have never been realized if it was not for the support of the whole ADL Team [1]. I am especially grateful to Philippe Ghez who was my very first early adopter and provided me with enough criticisms to fill up my agenda and support and praises to keep me actually working on the module. The best carrot & stick ever!

The entire ADL team has also been invaluable in all discussions concerning the ADLModule, its scope and purposes: they allowed me to stay focused rather then evolving the ADLModule into an mp3 tuner.

Finally, to all users which have reported bugs and suggestions: Thanks! You made me feel I was not working in a white tower.

## References

[1] A. Bazan, T. Bouedo, P.Ghez, M.Marino, C.E.Tull, "The Athena Data Dictionary and Description Language", 2003 Computing in High Energy and Nuclear Physics  (CHEP03), La Jolla, CA, USA, 2003

[2] "Athena, The ATLAS Common Framework", vs 2.0.2, August 2001, http://atlas.web.cern.ch/Atlas/GROUPS/SOFTWARE/OO/architecture/General/Tech.Doc/Manual/2.0.0-DRAFT/AthenaUserGuide.pdf

[3] A. Bazan, T. Bouedo, P.Ghez, C.E.Tull, "ADL Language Reference Manual", Release 1.0 – May 2002, http://atlas.web.cern.ch/Atlas/GROUPS/SOFTWARE/OO/architecture/DataDictionary/Documentation/Standard/Documents/index.html

[4] Together Control Center, Borland, http://www.togethersoft.com

[5] "Together for Atlas", http://atlas.web.cern.ch/Atlas/GROUPS/SOFTWARE/OO/tools/case/Together/

[6] S. Fisher, "Practical use of Together/Enterprise in ATLAS', http://atlas.web.cern.ch/Atlas/GROUPS/SOFTWARE/OO/tools/case/Together/use.html

[7] E. Gamma, R.Help, R.Johnson, J.Vlissides, "Design patterns: Elements of reusable object-orientedsoftware", Reading, MA, Addison-Wesley, 1995

**Insert PSN Here**